\documentclass{epl}

\usepackage{graphicx}
\usepackage{bm}
\usepackage{color}

\def\beq{\begin{equation}}
\def\eeq{\end{equation}}
\def\beqn{\begin{eqnarray}}
\def\eeqn{\end{eqnarray}}
\def\eqref#1{(\ref{#1})}

\def\vv#1{{\mathbf #1}}
\def\up{{\uparrow}}
\def\down{{\downarrow}}

\newcommand{\de}{\partial}

\def\pit{$(\pi/2,\pi/2)$}
\def\pio{$(\pi,0)$}

\newcommand{\myeq}[4]{\ifx#1#2#3\else#4\fi}

\def\pcite#1#2{\cite{#1}#2}

\title{
Weak phase separation and the pseudogap\\
in the electron-doped cuprates
}
\shorttitle{
Phase separation in electron-doped cupr.}

\author{M. Aichhorn and E. Arrigoni}

\institute{                    
Institut f\"ur Theoretische Physik - Computational Physics,
Technische Universit\"at Graz, Petersgasse 16, A-8010 Graz, Austria
}

\def\cha#1{{#1 }}

\begin{document}

\maketitle

\begin{abstract}

We study the quantum transition from an antiferromagnet to a
superconductor in a model for
electron- and hole-doped cuprates by means of a variational cluster
perturbation theory approach. In both cases, our results suggest a
tendency towards phase separation between a mixed
antiferromagnetic-superconducting phase at low doping and a pure
superconducting phase at larger doping.  However, in the
electron-doped case the energy scale for phase separation is an order
of magnitude smaller than for hole doping.
We argue that this can explain the different pseudogap and
superconducting transition scales in hole- and electron-doped
materials.

\end{abstract}

\section{Introduction}

High-temperature superconducting materials (HTSC) are characterized by
strong electronic correlations which are responsible for a number of
 anomalous  properties and competing phases.
The occurrence of these anomalous phases is probably related to the close
proximity of the Mott-insulating and antiferromagnetic (AF)  phases at half-filling.
This is particularly true for hole- (p-)doped materials which,
besides the 
AF and the superconducting (SC)
states, display  a number of unconventional phases such as
 stripes~\pcite{em.ki.99,ca.em.03}, checkerboard
structures~\pcite{ho.hu.02}, non-Fermi liquid phases, etc.
It has been argued that
these inhomogeneous phases
originate
 from 
an instability of the AF phase towards phase
separation~\pcite{ca.em.03},
which is intrinsic in
 models with short-range interactions such as Hubbard or
$t-J$ models.
On the other hand, long-range Coulomb repulsion enforces charge
homogeneity at long distances
so that stripes, other short-range charge inhomogeneities, or strong
charge fluctuations can originate as a
compromise between 
the two competing effects~\cite{ca.em.03,lo.em.94,ar.ha.02}. 
\cha{
A number of theories support the notion that
fluctuations of these inhomogeneities, or of related order parameters, are
 responsible for the pseudogap 
observed in p-doped materials~\cite{ca.em.03,ca.ca.99,sa.em.96}. In this case, 
 one would expect
 the corresponding (low) pseudogap temperature scale
$T^*$ to be indirectly related 
to the phase-separation energy scale. 
}

\cha{
The situation is not clear
in electron- (n-)doped cuprates (especially in the Nd- and Pr-CeCuO
compounds). These materials display a
more continuous transition from the AF to the SC
state~\cite{lu.le.90}. Stripes have not been observed so far, although
there are indications of electronic phase inhomogeneities 
~\cite{klauss,da.ka.05}.
In some mean-field calculations based upon the single-band Hubbard
model for n-doped materials, the AF gap 
closes continuously   leading to a
 quantum-critical point (QCP) and no instabilities towards inhomogeneous
phases are detected~\cite{ku.ma.02,se.la.05}, suggesting that the pseudogap is
directly related to fluctuations of the AF gap.
}
On the other hand, calculations based upon the three-band Hubbard model
yield stripes - although with different properties - even in n-doped
cuprates~\cite{sa.gr.00}. 

It is an important issue
 to understand 
whether the closing of the AF gap in n-doped cuprates, and thus
the evolution from the AF to the SC state may be
 related to an instability towards inhomogeneous phases,
in qualitative analogy to p-doped materials, or whether 
this evolution is more continuous, as the phase diagram of n-doped
cuprates seems to suggest~\cite{na.he}.

In this paper, we address the issue of the AF to SC transition 
by an analysis of the
single-band Hubbard
model for parameters appropriate to  n- and  p-doped compounds
via  {\em variational cluster perturbation theory}
(VCPT)~\pcite{po.ai.03,pott.03,da.ai.04}. VCPT  is appropriate to deal with
strongly-correlated systems, since the short-range interaction part is
solved exactly within a small cluster.
A similar  analysis has been recently carried out 
 by S{\'e}n{\'e}chal {\em et al.}~\pcite{se.la.05}. In that paper,
 the authors show that the single-band Hubbard model is sufficient to
 explain the different shape of the phase diagram between p- and n-doped
 cuprates. Their results also suggest that the AF to SC transition 
is continuous and associated with a quantum critical point.
{\em In contrast,} our results show
 that the issue of phase separation for the AF to SC transition 
in n-doped materials
{\em requires a high
accuracy} in the determination of energies, which
is possible within the present work only thanks to
two technical improvements of the VCPT calculations, which we are
going to describe in detail below.
\cha{
The interplay between AF and SC  was also studied previously via a
number of numerical techniques, see, e. g. Ref.~\cite{ga.ch.05}.
}

\begin{figure}[t]
 \includegraphics[width=\textwidth]{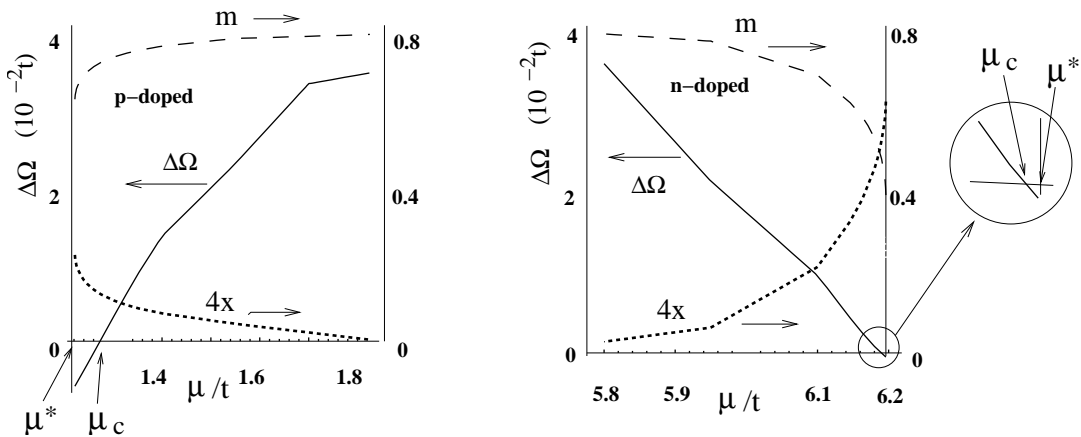}
\caption{
Difference between the $T=0$ grand-canonical potentials
 $\Delta \Omega\equiv
\Omega_{SC}-\Omega_{AF+SC}$
of  the SC and of the AF+SC
phases (solid line)
plotted as a function of chemical potential $\mu/t$.
 $m$ gives the  staggered magnetization (dashed),
 $x$ the doping (dotted, scaled by a factor $4$ for convenience) in the AF+SC phase. 
Data are for
the Hubbard model
with $U=8 t $ and $t'/t=-0.3$ for hole and electron doping.
Right: blowup of the region around the transition for the n-doped
case: Here,  the critical chemical potential $\mu_c$ at which
$\Delta\Omega =0$ is $\mu_c\approx 6.189$, and
$\mu^*\approx 6.196$ (see text for its meaning).
In both p-doped and n-doped system, $m$ is finite at the transition
point, and has a jump to $m=0$.
}
\label{vsmu}
\end{figure}

Our main results concerning the nature of the AF to SC transition
(see Fig.~\ref{vsmu})
are the following:
In  p-doped systems the situation is quite clear: the transition is
first order in $\mu$, i.e., there is an
instability towards 
 phase separation with a jump in doping $x$ and staggered
 magnetisation $m$ at the transition ($\mu_c$ in Fig.~\ref{vsmu}).
 We stress again that by allowing for a more general
 spatial dependences of the order parameter, this macroscopic phase
 separation could be possibly replaced by other microscopically inhomogeneous
 phases, such as, e.g., stripes. Certainly, this is expected to be the
 case if long-range Coulomb interaction is taken into account~\pcite{ca.em.03}.
On the other hand, in contrast to previous theoretical calculations,
our results
{\em also 
 suggest   phase separation  in the
n-doped case}, although  the corresponding  energy scale 
(see blowup in Fig.~\ref{vsmu}) 
is one order of
magnitude smaller than in p-doped compounds.

In agreement with  Ref.~\cite{se.la.05},  we find that the AF phase
actually mixes with
a weak d-wave SC component at small doping. 
A similar coexistence
phase 
is observed, for example, in PrCeCuO~\cite{da.ka.05}, and 
was obtained  in several
calculations~\pcite{li.ka.00, in.do.88}.
On the other hand, such a cohexistence obtained at the mean-field
level could be 
just a signal for the presence of
 strong pairing fluctuations within the
AF phase, as we discuss below.
\cha{
 Upon increasing doping, there is a
transition to a pure d-wave SC phase.
At larger doping,  
 some experiments and theories suggest a transition to  a SC phase with
 different  pairing symmetries in n-doped cuprates, such as
 $s$ or mixed $s$ and $d$ phases~\cite{st.kr.95,ch.jo.90,sk.ki.02}.
 However, this high-doping regime in beyond the scope of the present paper.
}

\section{Method}
We consider the two-dimensional single-band Hubbard model  with nearest-neighbor ($t$)
and next-nearest-neighbor ($t'$) hoppings, and Hubbard repulsion $U$.
We take typical parameters valid for both hole- and electron-doped high-$T_c$
cuprates~\pcite{ki.wh.98}, namely $t'/t = -0.3$ and $U/t=8$. 
\cha{
Although 
some slight difference have been suggested between the
parameter values 
($t'/t$ and $U/t$) of n- and p-doped compounds, we keep them equal in
order to {\em just focus on the differences due to hole- and electron-
  doping only}. 
We stress that
the aim of our calculation is a general 
description of the materials, 
so that we don't aim at carrying out an accurate fitting of the parameters.
We have in fact checked that 
different values for the parameters (e. g., a different $U/t$ or $t'/t$
within $30\%$) or 
a third-neighbor hopping  does not affect our 
qualitative
conclusions.
}

Within the CPT approach~\pcite{gr.va.93,se.pe.00}, 
the  lattice is partitioned into disconnected  clusters. The
Hamiltonian $H'_{\rm CPT}$ of the disconnected lattice can be solved
by exact diagonalization. The intercluster hopping Hamiltonian $V_{\rm CPT}$ 
is then
treated perturbatively up to first order for the self-energy.     
VCPT (or SFA)~\cite{pott.03,da.ai.04}
generalizes this approach 
by decomposing
 $H$
into a ``reference'' part $H'$ and a ``perturbation''
$V$, the latter containing single-particle terms only.
In general, one can take  $H'=H'_{\rm CPT}+\Delta H$, i. e., add to
the cluster an {\em arbitrary
  single-particle Hamiltonian}
  $\Delta H$, as long as $H'$ is still exactly solvable numerically. 
In order to conserve the total physical Hamiltonian
  $H=H'_{\rm CPT}+V_{\rm CPT}$, $\Delta H$ must then be
 subtracted from $V$~\pcite{da.ai.04}, which requires
$V=V_{\rm CPT}-\Delta H$.
This extension allows for the description of symmetry-broken phases by 
introducing, via $\Delta H$, a  corresponding symmetry-breaking field.
\cha{
Notice that,  due to this subtraction  the field $\Delta H$ is only fictitious.
 However, due to the fact that the subtraction is only carried out
in a perturbative way, results in fact do depend on $\Delta H$, except for the
case where CPT becomes exact.
}
This apparent ``arbitrariness'' in the choice of $\Delta H$ 
is restricted by the requirement that the
{\it self-energy  functional~\cite{pott.03}}
(which corresponds to the CPT grand-canonical potential)
\beq
\label{omega}
\Omega = \Omega'+ {\rm Tr} \ln (-G_{\rm CPT}) - {\rm Tr} \ln (-G') 
\eeq
has to be  stationary with respect
to $\Delta H$.
In \eqref{omega}, $\Omega'$ and $G'$ are the grand-canonical potential
and the Green's function  of the reference system 
$H'$, 
respectively, and $G_{\rm CPT}$ is
the Green's function of the physical system $H$ calculated
perturbatively in $V$~\cite{da.ai.04}. 
Physically, one can consider VCPT as an extension of CPT in which the
intercluster perturbation is not carried out with respect to the
exact cluster ground state but with respect to an
``optimized'' state parametrized by $\Delta H$. 
This is similar in
spirit to diagrammatic expansions
 in which the ``bare'' Green's
functions are replaced by ``dressed'' mean-field Green's functions in
some symmetry-broken state such as AF or SC.

Since we expect to describe both an AF and a d-wave SC phase,
in our calculation 
 $\Delta H$ contains both 
a staggered ($h$) and a nearest-neighbor d-wave pairing field
($\Delta$).
More specifically, 
$\Delta H=\Delta_{AF}+\Delta_{SC} + \Delta_{\epsilon}$,
where
$
\Delta_{AF} = h \sum_{\vv R} ( n_{\vv R ,\up} - n_{\vv R ,\down} ) e^{i \vv
  Q \cdot \vv R}
$,
and
$
\Delta_{SC}=
\frac{\Delta}{2} \sum_{\vv R,\vv R'} \eta(\vv R-\vv R') ( c_{\vv R ,\up} c_{\vv R'
  ,\down} + h.c.) 
$.
Here, $c_{\vv R ,\sigma}$ destroys an electron on lattice site $\vv R$
with spin projection $\sigma$.  $n_{\vv R ,\sigma} 
\equiv c^{\dag}_{\vv R,\sigma} c_{\vv R ,\sigma}$ is the corresponding density
operator, $\vv Q = (\pi,\pi)$ is the antiferromagnetic wave vector,
and the $d$-wave factor
$\eta(\vv R-\vv R')$ is non vanishing for nearest-neighbor lattice sites
only and is equal to 
 $+1$ ($-1$) for $\vv R-\vv R'$ in $x$ ($y$) direction.
In the SC term, the sum is restricted to $\vv R$ and $ \vv R'$
belonging to the same cluster.

Away from half-filling, in addition to the symmetry-breaking terms,
it is necessary to add to the reference system
a ``fictitious'' on-site 
energy, i. e.
\beq
\Delta_{\epsilon} = \epsilon \ \sum_{\vv R,\sigma }  n_{\vv R ,\sigma}
\;,
\eeq
which
plays the role of a ``shift'' in the cluster chemical potential
 with respect to the ``physical'' chemical potential $\mu$.
Without this term, which has been omitted in Ref.~\cite{se.la.05},
the mean particle density $n=1\mp x$ ($x$ is the doping, $\mp$
corresponds to p- and n-doping, respectively) cannot be unambiguously
determined, as different results would be obtained by evaluating it as
$n=-\de \Omega/\de \mu$ or as the usual trace over the Green's function.
Of course, a consistent treatment of the particle density is important
for an accurate analysis of phase transitions as a function of doping,
and of phase separation.
Unfortunately, the inclusion of $\epsilon$ considerably complicates
the numerical treatment, as now $\Omega$ is not simply a minimum with respect
to its parameters, but a saddle point (typically a maximum in the
$\epsilon$ direction).
Physically, one can regard $\epsilon$ as a Lagrange multiplier which
enforces an appropriate
constraint in the particle number~\pcite{details}. 
Along this
constraint, $\Omega $ is then a minimum with respect to $h$ and $\Delta$.

For an appropriate characterization of the phase transition, it is necessary
to evaluate $\Omega$ with high accuracy. This is difficult to achieve
 by simply numerically carrying out the frequency integrals 
contained in the trace in \eqref{omega}.
 It is rather
convenient to exploit the fact that these integrals can be replaced 
(at zero temperature $T$)
by a sum over the negative poles 
of
$G_{\rm CPT}(\vv q)$ and
of
$G'$ (see Ref.~\cite{pott.03.se} for details).
While the latter are directly given by the
single-particle and single-hole 
excitation energies 
 of the cluster, the former
can be directly obtained as the eigenvalues 
of the
matrix $h_{\alpha,\beta}(a,b)$ in Eq. (17) of Ref.~\cite{za.ed.02}~\pcite{details}.
Since this method requires the evaluation of
 all cluster single-particle and single-hole states,
 a complete diagonalization of the cluster Hamiltonian must
be carried out. In addition, the dimension of the matrix $h_{\alpha,\beta}(a,b)$
is equal to the number of  single-particle and single-hole states.
This
 restricts the maximum cluster sizes that can be
considered in our VCPT calculation.
For this reason, we only consider an infinite 
 lattice split into  $2\times2$ clusters as a
reference system.
We stress again that this is necessary
in order to achieve an accuracy that
permits the resolution of the tiny phase-separation energy scale in
the n-doped case (see blowup in Fig.~\ref{vsmu}).
\cha{
Of course one could argue that finite-size corrections may introduce
a even larger inaccuracy. However, while quantitative changes are
expected for larger clusters, 
we observe that our result about phase separation is
quite robust and only weakly dependent on other parameters such as 
$t'/t$ for  $0.2<|t'/t|<0.4$, or $U/t$. In addition, our calculation {\em does} show
that {\it at least} such an accuracy is required  
in order to address the issue of phase
separation in the n-doped case.
}

\section{Results}
Our calculation, thus, proceeds as follows. For a given value of the
variational parameters $\epsilon$, $h$, and $\Delta$ we carry out a
complete diagonalization of 
the cluster Hamiltonian (due to the presence of the pairing field,
particle number is not conserved),  evaluate $\Omega$ 
as described above,
and look for a stationary solution as a function of
the variational parameters.
Quite generally,  different stationary solutions can be found,
corresponding, for example, to different phases. In this case, the
minimum $\Omega$ selects the most stable phase.
Near half-filling, the two most stable solutions are a coherently mixed AF+SC
and a pure SC phase, in agreement with Ref.~\cite{se.la.05}.
 Results for these solutions are plotted in
Fig.~\ref{vsmu} for the p- and n-doped cases.
 The AF order parameter $m$ is plotted as a dashed line
for both n- and p-doped systems as a function of $\mu$.
$m$ monotonously decreases upon going away
from half-filling.
In Fig.~\ref{vsmu}, we also plot the doping $x$ and the difference
between the $T=0$ grand-canonical potentials
 $\delta \Omega\equiv
\Omega_{SC}-\Omega_{AF+SC}$
 of 
the SC and the AF+SC
 solutions.
While at lower doping the AF+SC phase is stable,
 a crossing to the SC phase
 occurs at a critical chemical potential $\mu_c$ for which $m$ is
 still nonzero, { i. e. the transition is first order as a function of $\mu$ 
\em both for the p- as well as for the n-doped case}, as anticipated.
Although, from Fig.~\ref{vsmu}, this effect appears to be very small for the n-doped
case,  we stress that our calculation was carried out with sufficient
 accuracy to resolve this energy difference.
At this value of $\mu$, the doping $x_c$ of  the AF+SC phase is
different from the one $x_c'$ of 
the pure SC
phase so 
that there is a jump $\Delta x \equiv x_c'-x_c>0$ in the doping at $\mu_c$ indicating
phase separation between a weakly doped AF+SC  and a higher doped SC
phase.
Eventually,  for some $\mu=\mu^*$ (further away from half filling than $\mu_c$)
 the AF+SC solution ceases to exist.
In principle, an unstable AF+SC solution continues to exist for 
$x>x_c$. However, in this region $\mu$ is a decreasing function of the particle density,
which is an equivalent indication of phase separation. 

The difference $\Delta\mu \equiv|\mu^* -\mu_c|$ can be considered as a
measure of the characteristic
phase-separation energy per particle, i.e., it is proportional to the energy
barrier 
between the two doping values $x_c$ and $x_c'$.
As one can see from Fig.~\ref{vsmu}, 
in the p-doped case this energy scale 
($\Delta\mu/t \approx 5 \cdot \ 10^{-2}$) 
is about an order of magnitude
larger than in the n-doped case 
($\Delta\mu/t \approx 7 \cdot \ 10^{-3}$). 
Taking typical values for the energy unity $t$ ($t\approx 0.25 eV$),
 this corresponds to a
temperature scale
of $\sim 125 K$ in p-doped,
and $\sim 18 K $ in electron- doped cuprates, which 
is roughly of the order of the
corresponding (low-energy) pseudogap temperature scales observed in
these materials~\cite{al.kr.03}.
Also the discontinuity 
$\Delta x$ (not shown)
is larger in the p-doped case ($\Delta x = 0.11 $, vs. $\Delta x =
0.08 $ for n-doped), although
the difference is not significant.
At the same time, the doping  at which the AF+SC solution is destroyed
 is much
larger in n-doped ($x_c\approx 0.13\%$) than in p-doped systems
($x_c\approx 0.03\%$), in qualitative agreement with
experiments~\cite{lu.le.90,na.he}. 
Therefore, our results suggest that the AF+SC to SC transition as a
function of $\mu$ is clearly
first order in the p-doped case while weakly first order
for n-doping. 

\section{Single-particle spectrum}
\begin{figure}[t]
  \includegraphics[width=\textwidth]{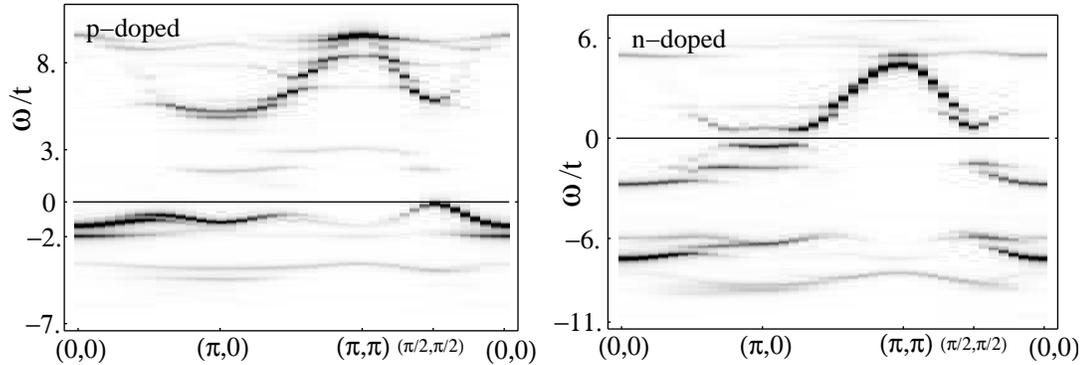}
\caption{
Single-particle spectrum for the hole-doped   (a), and for the electron-doped
system  (b)  just before the AF+SC to SC transition, i.e., for
dopings 
$x \approx 0.03$ and $x \approx 0.13$, respectively.
}
\label{spectr}
\end{figure}

The different behavior between n- and p-doped  systems can be understood
by  considering the
 doping evolution of the single-particle spectrum from the
AF insulator to a SC displayed in Fig.~\ref{spectr}.
In agreement with
experiments~\pcite{ar.ro.02,ar.lu.01},  in p-doped cuprates doped holes
  first enter at \pit{}\ \pcite{da.hu.03}.
On the other hand, in n-doped materials doped electrons 
 initially form pockets around \pio~\pcite{ar.ro.02,ar.lu.01,ku.ma.02}.
At \pio the density of states is larger, and the fact that 
$\mu$ lies within the SC gap apparently 
  stabilizes the AF solution for a larger doping range, allowing for
  the AF gap to  decrease more gradually in the n-doped case. The transition
  to a non-magnetic solution appears to occur as soon as the Fermi
  energy touches the top of the band at \pit.
At this nodal point the SC gap is zero, so that doping into nodal
particles apparently destabilizes the AF solution.
This 
appears to be the
reason 
why
in the p-doped case
 the AF solution, 
 where hole are first doped near \pit (see also Ref.~\cite{ku.ma.02}),
is stable only in a smaller doping range.
 Numerically, we observe the following behavior:
 as long as $\mu$ remains below (above for p-doped) the band around \pit, there is
 only an absolute
  minimum of $\Omega$ 
 at a finite value $h_{opt}$ of the 
 staggered
 field  $h$ (we keep the other two fields $\Delta$ and $\epsilon$ at
 their saddle point).
 As soon as $\mu$ enters
  the band at \pit,  $\Omega(h)$ also develops a local minimum at
   $h=0$, 
  which becomes rapidly lower in energy than the
 minimum  $h=h_{opt}$. Eventually, the local maximum lying between the two minima at
 $h=0$ and $h=h_{opt}$ merges into the minimum at 
 $h=h_{opt}$ so that this latter minimum disappears.
The observation that doping into \pit makes the AF phase unstable 
suggests that the occurrence of phase separation is generic and quite 
independent on specific parameters (unless, of course, one adds some
doping-dependent potential).
Finally, 
let us mention that the Fermi-surface evolution of the
n-doped system as a function
of $x$ (not shown) qualitatively reproduces the ARPES
experiments~\cite{ar.lu.01}:
For small doping, we obtain electron pockets around \pio, while for
larger doping our results display an evolution to a large 
 Fermi surface centered around $(\pi,\pi)$.

\section{Discussion and conclusions}
The VCPT method exactly treats  fluctuations up to the range of the cluster
size, so that 
the question arises whether the SC solution we (and also
others~\cite{ch.jo.90,li.ka.00,se.la.05}) 
  obtain within the AF phase is a
true long-range SC phase or whether it is only a signal of  strong pairing fluctuations
within the AF phase leading to a SC pseudogap.
\cha{
The latter hypothesis could be supported by the fact that 
 results obtained with different  cluster
sizes~\pcite{se.la.05} 
seem to indicate a  size dependence of the SC order parameter, and by
the fact that the SC order parameter is about a factor three smaller in the AF+SC
phase than in the pure SC one.
The presence or not of such microscopic cohexistence phase may depend 
on material details.
Certainly, our results suggest that the
 SC gap (or pseudogap in the case of fluctuations) $\Delta$
is important in order to stabilize the AF phase in n-doped materials.
On the other hand, $\Delta$  hardly has an effect in p-doped systems
since it does not produce any gap at the Fermi Surface,
 as doped holes first enter near the nodal points.
}

In conclusion, our calculations suggest that 
the destruction of the AF phase  in n-doped systems is associated with a
tendency towards phase separation, similarly to  p-doped materials, although
with an energy scale which is one order of magnitude smaller.
 In the presence of Coulomb interaction,
this tendency is expected to produce  a microscopically segregated phase,
such as stripes or similar inhomogeneous structures~\cite{sa.gr.00}
Following the idea that
the pseudogap temperature is related to the formation of such
 segregated phases~\cite{ca.em.03}, our results 
provide reasonable orders of magnitudes for the pseudogap temperature scale, and,
in particular, provide a possible explanation why
this is much smaller in n-doped than in p-doped
cuprates.
\cha{
The results of the present paper 
support the idea 
that  the asymmetry between hole- and
electron-doped materials can be understood within a simple {\em single-band
Hubbard model}, at least 
concerning the portion of phase diagram we
studied here. Of course, it remains an open issue, whether 
a three-band model may be necessary for a complete description.
In fact, alternative theories propose that Ce- doping 
in n-doped compounds,  such as NdCeCuO, 
occurs in a total different way. The idea there is that doped Ce atoms do not add
electrons to the CuO$_2$-layers, but rather are responsible for a
quenching of  Cu moments~\cite{bask.05u}.
}

\acknowledgments
We thank M. Potthoff and L. Alff for discussions and
useful suggestions.
This work is partially supported by the Doctoral Scholarship 
Program of the Austrian Academy of Sciences (M.A.) and by the 
DFG {\it Forschergruppe} n. 538.

\bibliographystyle{prsty-etal} 
\bibliography{footnotes,references_database}

\end{document}